\documentclass[preprint2,times]{aastex62}
\usepackage{amsmath}

\shorttitle{slow waves in a multi-stranded coronal loop}
\shortauthors{Krishna Prasad et al.}

\begin{document}
\title{Modelling the propagation of slow magneto-acoustic waves in a multi-stranded coronal loop}

\correspondingauthor{S. Krishna Prasad}
\email{krishna.prasad@aries.res.in}

\author[0000-0002-0735-4501]{S. Krishna Prasad} 
\affiliation{Aryabhatta Research Institute of Observational Sciences, Nainital-263001, India}                              
\author{T. Van Doorsselaere}
\affiliation{Centre for mathematical Plasma Astrophysics, Mathematics department, KU Leuven,
Celestijnenlaan 200B, 3001 Leuven, Belgium}                                  

\begin{abstract}
We study the propagation properties of slow magneto-acoustic waves in a multi-thermal coronal loop using a 3D MHD model, for the first time. A bundle of 33 vertical cylinders, each of 100{\,}km radius, randomly distributed over a circular region of radius 1{\,}Mm is considered to represent the coronal loop. The slow waves are driven by perturbing the vertical velocity ($v_z$) at the base of the loop. We apply forward modelling to the simulation results to generate synthetic images in the coronal channels of SDO/AIA. Furthermore, we add appropriate data noise to enable direct comparison with the real observations. It is found that the synthetic images at the instrument resolution show non-cospatial features in different temperature channels in agreement with previous observations. Time-distance maps are constructed from the synthetic data to study the propagation properties. The results indicate that the oscillations are only visible in specific channels depending on the temperature range of plasma existing within the loop. Additionally, the propagation speed of slow waves is also found to be sensitive to the available temperature range. Overall, we propose that the cross-field thermal properties of coronal structures can be inferred using a combination of numerical simulations and observations of slow magneto-acoustic waves.

\end{abstract}


\section{Introduction}
There is ample evidence in the literature suggesting that a sizable fraction of the solar coronal loops possess multithermal cross-sections. Because the cross-field thermal conduction is heavily suppressed in the solar corona, such structures can exist as multi-stranded loops. The volume filling factor for the plasma within the coronal loops is estimated to be a few to several tens of \%, indicating the presence of fine structure within the coronal loops \citep{1999A&A...342..563D, 2008ApJ...686L.131W, 2009ApJ...694.1256T, 2015ApJ...800..140G}. Similar characteristics were found in active region fan loops \citep{2012ApJ...744...14Y}. 

It has been observed that coronal structures appear diffuse (or fuzzy) at hot temperatures and sharp at warm temperatures \citep{2007A&A...466..347D, 2009ApJ...694.1256T}. Eclipse observations in the past also displayed similar behaviour \citep[see for e.g., Fig.{\,}2 of][]{2006SoPh..236..245S}. An explanation was provided in terms of sub-resolution strands within coronal loops \citep{2010ApJ...719..576G}, but it necessitated the reemergence of sharper structures at very hot ($\sim$6{\,}MK) temperatures. Later, \citet{2011ApJ...736L..16R} showed that thin, discrete structures indeed reappear at hotter temperatures and thus confirmed the existence of sub-resolution structures within coronal loops. 

Using high-resolution spectroscopic observations in H$\alpha$, \citet{2012ApJ...745..152A} studied the dynamics and morphological properties of coronal rain. The authors find that the downfalling plasma condensation blobs roughly trace the coronal loops but have much narrower widths (on an average $\sim$ 310{\,}km) suggesting the presence of fine structure within those loops. \citet{2012ApJ...750L..25J} employ high-resolution imaging observations in He{\,}\textsc{i} 10830{\,}{\AA} line, to identify ultrafine magnetic structures of 100{\,}km diameter cospatial to an active region loop system. The observed magnetic braiding \citep{2013Natur.493..501C} and nano jets \citep{2021NatAs...5...54A, 2022ApJ...938..122P} also support sub-resolution structuring within coronal loops. Other studies investigated the distributions of cross-sectional widths of coronal loops and additionally compared them across different instruments to identify if the structures are completely resolved \citep{2017ApJ...840....4A, 2020ApJ...902...90W}.

The cross-field thermal structure of a coronal loop is generally obtained from multi-wavelength data using techniques such as Emission Measure (EM) loci \citep[e.g.,][]{2002A&A...385..968D} and Differential Emission Measure (DEM) \citep[e.g.,][]{2012A&A...539A.146H}. Early studies produced contradicting results with some researchers suggesting that coronal loops have an isothermal cross-section \citep{2003A&A...406.1089D, 2007ApJ...655..598C, 2008ApJ...674.1191N, 2009ApJ...694.1256T} while some others indicate that they instead have a multi-thermal cross-section \citep{2002ApJ...577L.115M, 2005ApJ...627L..81S, 2006ApJ...636L..49S, 2007ApJ...667..591P}. Additionally, \citet{2008ApJ...686L.131W} found coronal loops which are not entirely isothermal but have a narrow temperature distribution over their cross-sections. Later studies involving the analysis of multiple loop structures show the existence of both isothermal and multi-thermal loops \citep{2011ApJ...732...81A, 2012ApJ...755L..33B, 2016ApJ...831..199S}.

The multi-stranded and multithermal nature of coronal loops is also evident from the observations of waves and oscillations. For instance, the observations of transverse decayless oscillations often show a coherent wavy pattern across the multiple threads of a loop \citep{2012ApJ...751L..27W, 2013A&A...552A..57N, 2021A&A...652L...3M}. Furthermore, the different dynamical behaviour of these oscillations across multiple wavelengths also suggests a multi-thermal nature of the loop. It is believed that the differential propagation of slow magnetoacoustic waves, as observed using multiple wavelength channels, indicates a multithermal structure within a loop \citep{2003A&A...404L...1K}. In fact, \citet{2017ApJ...834..103K} extracted two individual components of a multithermal loop using the observations of propagating slow waves. Additionally, \citet{2021ApJ...914...81K} have shown that the multi-wavelength behaviour of standing slow waves also suggests the presence of multithermal structure within a coronal loop.

Coronal loops with multiple strands were also considered in numerical models by a number of authors \citep[e.g.,][]{2009ApJ...694..502O, 2010ApJ...714.1239K, 2010ApJ...716.1371L, 2016ApJ...823...82M, 2019ApJ...883...20G}. In this article, we study the propagation of slow magnetoacoustic waves in a coronal loop using a multi-stranded and multithermal loop model for the first time. In such a model, because the slow waves propagate at different speeds in different strands owing to their differences in temperatures, they gradually go out of phase, resulting in the cumulative oscillation amplitude decreasing over a distance. This implies the slow waves propagating in a multithermal loop will appear to be damping even though there is no real dissipation of energy. As we will see, this damping indeed occurs in our model. We call this the Multithermal Apparent Damping (MAD) which is analysed separately in detail in \citet{2024arXiv240109803V}. Concurrently, a similar damping effect is seen by \citet{2024SoPh..299....2F}, who modelled the propagation of slow magnetoacoustic waves in a plasma slab with smoothly varying temperature and density profiles transverse to the magnetic field. We describe our numerical setup, including the driver details, in Section{\,}\ref{simul}, followed by a discussion of obtained results in Section{\,}\ref{res}, and at last we present our conclusions in Section{\,}\ref{concl}.

\section{Numerical simulations}
\label{simul}
We perform 3D MHD simulations to study the propagation of slow magneto-acoustic waves in a multi-thermal coronal loop. Considering ideal conditions (i.e., no dissipation terms), the MHD equations are numerically solved using the publicly available MPI-AMRVAC code \citep[version 2.0][]{2014ApJS..214....4P, 2018ApJS..234...30X}. We use a second-order predictor-corrector type two-step scheme for the time integration and the `hlld' flux scheme for the spatial discretization. The errors on the divergence of the magnetic field are treated using a Generalized Lagrange Multiplier (GLM) method. \\

\begin{figure*}
\centering
\includegraphics[scale=0.75]{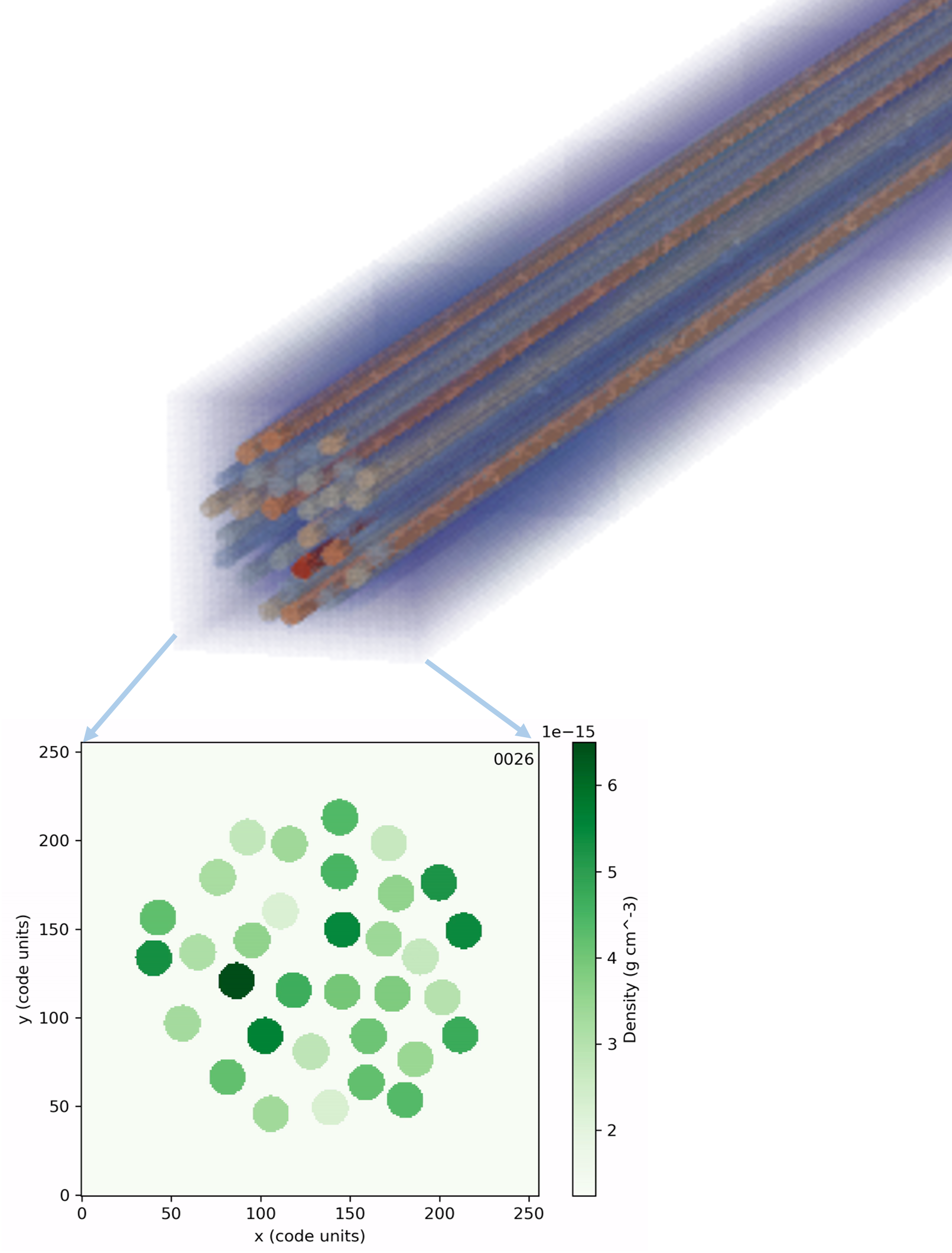} 
\caption{A 3D rendering of the multi-stranded loop model employed. The bottom panel shows the cross-sectional density profile highlighting the distribution of strands across the loop cross section.}
\label{fig1}
\end{figure*}

The numerical setup involves a bundle of vertical cylinders randomly distributed within a circular region (see Fig.{\,}\ref{fig1}) representing a multi-stranded loop. The radius of each strand is 100{\,}km and that of the loop is 1{\,}Mm. A total of 33 strands are considered resulting in a filling factor of 0.33 for the loop \citep{2005ApJ...621..498K, 2005A&A...439..351U, 2014ApJ...795...18V, 2015ApJ...800..140G}. The physical size of the simulation domain is considered to be 100{\,}Mm along the loop length ($z$-axis) and 2.5$\times$2.5{\,}Mm$^{2}$ in the perpendicular direction ($x$--$y$ plane). The corresponding number of grid cells is 128 along the $z$-axis and 256$\times$256 in the $x$--$y$ plane. Consequently, the spatial scale is $\approx$0.78{\,}Mm along the loop length and $\approx$9.76{\,}km in the cross-sectional plane. The plasma temperature within each strand is obtained from a random normal distribution whose peak is at log{\,}$T= 6.0$, and has a standard width of 0.15 in the logarithmic scale. The corresponding plasma (number) densities are also obtained from an uncorrelated random normal distribution with a peak value of log{\,}$n= 9.2$ and a standard width of 0.10. The resultant plasma temperature in the loop strands varies from 0.41{\,}MK to 1.73{\,}MK. Accordingly, the number densities vary from 1.2$\times$10$^{9}${\,}cm$^{-3}$ to 2.5$\times$10$^9${\,}cm$^{-3}$. These values fall in the typically observed range for coronal loops. The external (including those locations within the loop but outside the strands) plasma temperature and number densities are kept fixed at log{\,}$T= 6.0$ and log{\,}$n \approx 8.7$ (1/3rd of the internal peak value), respectively. The gravitational stratification is ignored, meaning the plasma parameters are the same all along the loop length. Also, following the definition of the strand, the plasma parameters are kept constant across each strand. The magnetic field $B$ is considered to be vertical ($B_x$,$B_y$=0), parallel to the loop axis, with a magnitude of 50{\,}G ($B_z$=50). We understand that the coronal magnetic field strength values typically inferred from the direct \citep{2000ApJ...541L..83L, 2004ApJ...613L.177L, 2019ApJ...874..126K, 2021ApJ...915L..24B} and indirect \citep{2001A&A...372L..53N, 2007Sci...317.1192T, 2016NatPh..12..179J, 2020Sci...369..694Y} observational techniques range from a few Gauss to a few tens of Gauss, but considering the complicated setup a strong field is required to keep the loop stable, so we consider a value on the higher side in this model. Moreover, since we are interested in only slow magneto-acoustic waves here, the effect of the magnetic field on their propagation is going to be negligible. The magnetic field is largely uniform across the domain except within the loop strands. In order to compensate for the high gas pressure within the hot and dense strands, the magnetic field is reduced so that the total pressure is uniform keeping the loop stable. However, the strong magnetic field also ensures the plasma beta is very small, and consequently, the reduction in the magnetic field required is marginal. The $B_z$ value across the domain varies from 49.7 -- 50 G.  \\

All the boundaries are kept open with zero gradients for all variables except for the vertical velocity ($v_z$) at the bottom boundary which we perturb to generate slow magneto-acoustic waves in the loop. The $v_z$ at the bottom of the loop is continuously perturbed using a sinusoidal driver of the form
\begin{equation}
v_z = v_{z0}{\,}\mathrm{sin}(2 \pi t/P).
\end{equation}
Here, $t$ is time, $P$ is the oscillation period, and $v_{z0}$ is the oscillation amplitude. The oscillation period is chosen as 180{\,}s, a typical value for propagating slow magneto-acoustic waves observed in active region loops. The oscillation amplitude is set to $\approx$7.6{\,}km{\,}s$^{-1}$ within the loop (0 outside the loop) which is about 5{\,}\% of the sound speed at the peak plasma temperature (log{\,}$T= 6.0$). This small amplitude keeps the generated waves in the linear regime and is in line with that found in observations \citep[for e.g.,][]{2012A&A...546A..50K,2017ApJ...850..206S,2019FrASS...6...57S, 2020ARA&A..58..441N}. The simulation is run for 1200 seconds in physical time which should approximately capture about seven oscillation cycles near the base of the loop.

\section{Results and Discussion}
\label{res}
The evolution of plasma density and velocity within the simulation clearly shows fluctuations corresponding to slow magneto-acoustic waves propagating along the individual strands within the loop. In order to compare the propagation properties of these waves with those from observations, we generate synthetic images using the forward modelling code FoMo \citep{2016FrASS...3....4V}. This code takes the plasma temperature, density, and velocity at each voxel (volume element) along a desired line-of-sight direction, and estimates the expected emissivities (taking solar abundances into account), which are subsequently convolved with the telescope characteristics (e.g., spectral transmission, detector sensitivity, etc.) and integrated along the line of sight, to synthesise intensity images for a particular wavelength channel/filter. Alternatively, one could also use the code to generate synthetic spectral lines following a similar procedure. Here, we employ the code to generate synthetic images for all the six coronal channels (94{\,}{\AA}, 131{\,}{\AA}, 171{\,}{\AA}, 193{\,}{\AA}, 211{\,}{\AA}, and 335{\,}{\AA}) of the SDO/AIA \citep{2012SoPh..275...17L, 2012SoPh..275....3P} assuming the line of sight is along the $y$-axis (i.e., perpendicular to the loop length). 

\begin{figure*}
\centering
\includegraphics[scale=0.90]{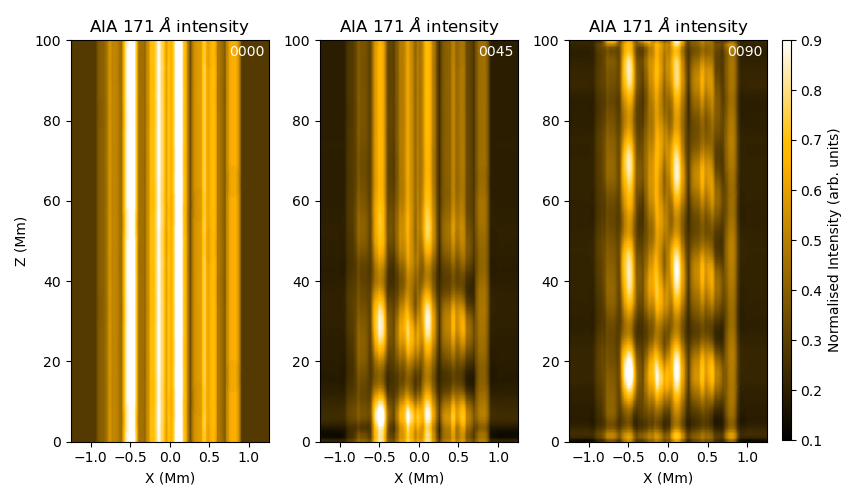} 
\caption{Sample synthetic images of the loop in AIA 171{\,}{\AA} channel generated assuming the line-of-sight direction to be perpendicular to the loop axis (i.e., along $y$-axis). The simulation resolution is still preserved in these images. The left, center, and right panels display the first, 45th and 90th frames corresponding to 0 s, 450 s, and 900 s, from the beginning of the simulation. The center and right panels show the compressive signature of propagating slow waves within individual strands. A movie associated with this figure showing the propagation of slow waves across the entire simulation duration (1200 s) is available online.}
\label{fig2}
\end{figure*}

\begin{figure*}
\centering
\includegraphics[scale=0.90]{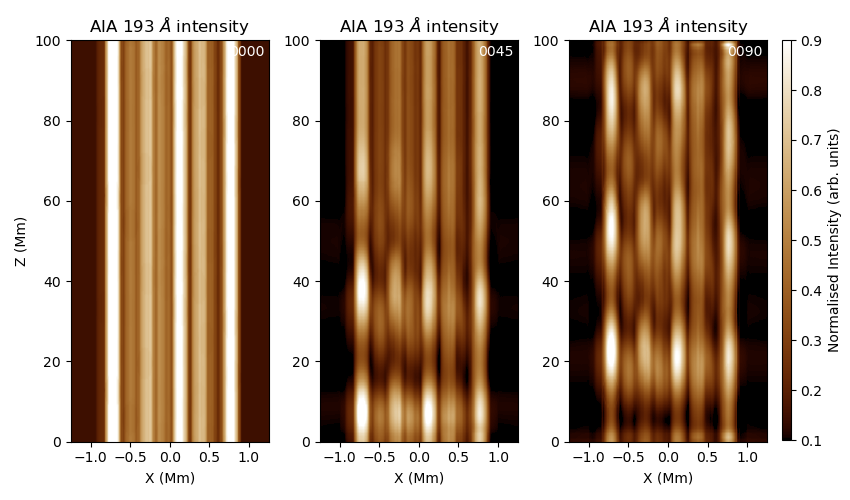} 
\caption{Sample synthetic images of the loop in AIA 193{\,}{\AA} channel generated assuming the line-of-sight direction to be perpendicular to the loop axis (i.e., along $y$-axis). The simulation resolution is still preserved in these images. The left, center, and right panels display the first, 45th and 90th frames corresponding to 0 s, 450 s, and 900 s, from the beginning of the simulation. The center and right panels show the compressive signature of propagating slow waves within individual strands. A movie associated with this figure showing the propagation of slow waves across the entire simulation duration (1200 s) is available online.}
\label{fig3}
\end{figure*}

\begin{figure*}
\centering
\includegraphics[scale=0.90]{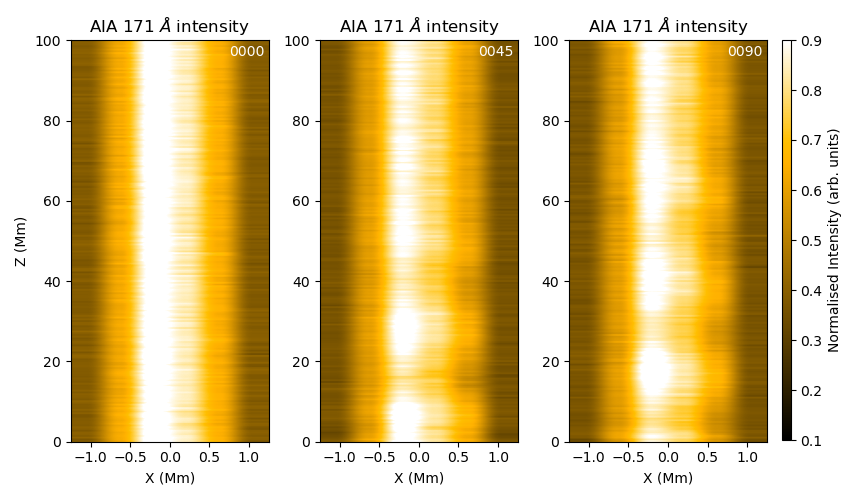} 
\caption{Sample synthetic images of the loop in AIA 171{\,}{\AA} channel after degrading the pixel scale to match with that of the AIA and adding corresponding data noise. The images in individual panels correspond to the same time frame as that shown in Fig.{\,}\ref{fig2}. Note that these images are quite a bit stretched in the $x$-direction to highlight the key differences.}
\label{fig4}
\end{figure*}

\begin{figure*}
\centering
\includegraphics[scale=0.90]{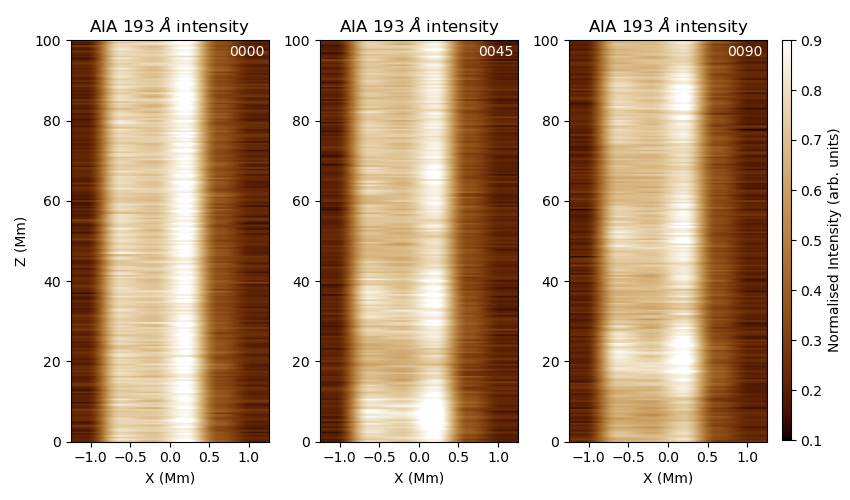} 
\caption{Sample synthetic images of the loop in AIA 193{\,}{\AA} channel after degrading the pixel scale to match with that of the AIA and adding corresponding data noise. The images in individual panels correspond to the same time frame as that shown in Fig.{\,}\ref{fig3}. Note that these images are quite a bit stretched in the $x$-direction to highlight the key differences.}
\label{fig5}
\end{figure*}

Sample images thus produced in the AIA 171{\,}{\AA} and 193{\,}{\AA} channels are shown in Figs.{\,}\ref{fig2} \& \ref{fig3}, respectively. The associated movies (available online) reveal how the simulated slow waves are manifested in these two wavelength channels. Notably, the compressive signature of slow waves is also visible in the centre and right panels of the images. Importantly, the cross-field structure of the loop appears different in these two channels. This is because these two channels are meant to observe different temperature plasma. Note that the resolution of the simulation is still preserved in these images. More particularly, the spatial scale in the transverse direction ($\approx$9.76 km per pixel) is much lower than that is actually achievable in SDO/AIA observations ($\approx$430{\,}km). Therefore, the intensities were rebinned to match the pixel scale to that of AIA. Along the $z$-axis, this would require a bit of interpolation, as the pixel scale in this direction is slightly larger ($\approx$780{\,}km) than that of AIA. Additionally, we also add a random noise component to these images that is generally expected in observations. The amplitude of the noise within each AIA channel is estimated by incorporating various components including the photon noise, readout noise, digit noise, compression noise, dark noise, subtraction noise, and the despiking noise \citep{2012A&A...543A...9Y, 2012SoPh..275...41B}. The resultant images from AIA 171{\,}{\AA} and 193{\,}{\AA} channels are shown in Figs.{\,}\ref{fig4} \& \ref{fig5} respectively.

Note that, after scaling them to AIA resolution, there are hardly 6 pixels in the transverse direction, so the images look quite stretched along the $x$-axis. Also, to avoid a jagged appearance due to the availability of very few pixels, the images in Fig.{\,}\ref{fig3} are smoothed using a bilinear interpolation. While this did not have any impact on the overall structure, it brings out a noisy appearance in the horizontal direction. Nevertheless, what appeared as multiple strands in Figs.{\,}\ref{fig2} \& \ref{fig3} seems to be flattened out in these images. Importantly, in the AIA 171{\,}{\AA} channel (see Fig.{\,}\ref{fig4}), the bundle of strands emerge as a single isolated structure which is what we regularly see in observations. The AIA 193{\,}{\AA} channel on the other hand, displays a bifurcated structure (see Fig.{\,}\ref{fig5}) with two strands which are seemingly located adjacent to the structure visible in the 171{\,}{\AA} channel. This is again because the two channels are capturing different temperature plasma. It may be noted that in the solar coronal observations, loops captured in different temperature channels are occasionally found next to each other rather than being exactly cospatial \citep{2003ApJ...590.1095N, 2004ApJ...601..530S, 2011ApJ...734..120H, 2014A&A...570A..84N}. Those loops are generally interpreted as multi-thermal loops. The natural reproduction of this behaviour in our simulations implies that one could indeed use this trait to infer the underlying multi-thermal nature of coronal loops. \\

\begin{figure*}
\centering
\includegraphics[scale=0.65]{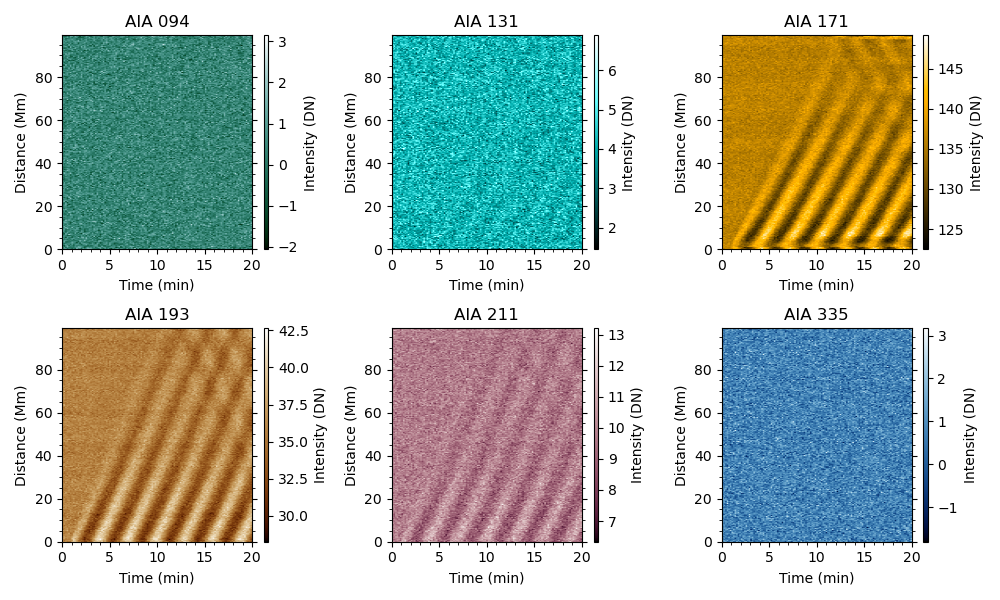} 
\caption{Time-distance maps constructed from the synthetic data for all the six coronal channels of AIA. The slanted ridges of alternating brightness visible in some of the channels highlight the propagation of slow waves along the loop.}
\label{fig6}
\end{figure*}

Subsequently, we collapse the loop in the transverse direction and stack the intensity profiles from each time step next to each other to build a time-distance map. Such maps for all the AIA channels are displayed in Fig.{\,}\ref{fig6}.
As can be seen, slanted ridges of alternating brightness, a signature of the propagating slow magneto-acoustic waves, are visible in the AIA 171{\,}{\AA}, 193{\,}{\AA}, and 211{\,}{\AA} maps. The temporal spacing between the ridges is about 3 minutes which is the same as the period of the driver. Additionally, in these maps, the backward slanted ridges may be seen near the top of the domain (forming a crisscross pattern) indicating some unwanted reflection from the top boundary. But, we have data covering enough spatial extent to study the propagation properties so the reflection part can be ignored for the present study. More importantly, the ridges are very faint in the AIA 131{\,}{\AA} channel and not at all visible in the 94{\,}{\AA} and 335{\,}{\AA} channels. This behaviour is also seen in observations and the specific channels the ridges are visible will apparently vary from loop to loop. For instance, the active region loop studied by \citet{2012ApJ...757..160J} shows propagating waves in all channels except the 94{\,}{\AA} while that by \citet{2020A&A...638A...6S} does not show any oscillations in the 335{\,}{\AA} and 94{\,}{\AA} channels. A few other studies report oscillations in fewer channels often because of the lower amplitudes at remaining wavelengths \citep{2012SoPh..279..427K, 2013ApJ...778...26U, 2017ApJ...834..103K}. In the current case, the absence of oscillations in some of the channels is because the loop does not possess plasma at their respective temperatures. \\

The characteristic peak response temperatures for the AIA 94{\,}{\AA}, 131{\,}{\AA}, 171{\,}{\AA}, 193{\,}{\AA}, 211{\,}{\AA}, and 335{\,}{\AA} channels are 7.08{\,}MK, 0.56{\,}MK, 0.79{\,}MK, 1.41{\,}MK, 1.78{\,}MK, and 0.22{\,}MK respectively. The 94{\,}{\AA}, 131{\,}{\AA}, and 335{\,}{\AA} channels have additional peaks at 1.58{\,}MK, 11.2{\,}MK, and 2.5{\,}MK respectively. These values are obtained from peak locations in the AIA temperature response curves \citep[version 10;][]{2014SoPh..289.2377B}. Since our simulated loop has plasma temperature varying between 0.41{\,}MK to 1.73{\,}MK, the oscillations are expected to be visible in the 131{\,}{\AA}, 171{\,}{\AA}, and 193{\,}{\AA} channels. However, the oscillations are nearly invisible in the 131{\,}{\AA} channel and on the other hand we see oscillations in the 211{\,}{\AA} channel albeit at a lower amplitude. This is perhaps due to the differences in the number of strands and densities at the extreme temperatures and could also be partly related to the sensitivities of the respective detectors. Nevertheless, this behaviour highlights the importance of forward modelling taking all the instrumental effects into account and clearly indicates that one cannot simply conjecture the cross-field thermal information based on the peak response temperatures. The lower amplitudes for slow waves, observed in the hot 211{\,}{\AA} channel, are generally attributed to high thermal conduction in hotter plasma. However, in the current setup, there are no dissipative mechanisms and there is no plasma corresponding to the peak response temperatures of 211{\,}{\AA}. Therefore, one should exercise caution in directly attributing the low (or high) oscillation amplitudes to the strength of dissipative processes. \\

Lastly, we compute the propagation speeds of slow waves observed in the 171{\,}{\AA}, 193{\,}{\AA}, and 211{\,}{\AA} channels. To achieve this, the light curve at each spatial location is cross-correlated with that at the adjacent position and the corresponding time lags are measured as a function of distance along the loop. For this purpose, distances between 4 -- 48{\,} Mm are only considered to avoid boundary effects at the bottom (see 171{\,}{\AA} image) and reflections at the top.
\begin{figure*}
\centering
\includegraphics[scale=0.75]{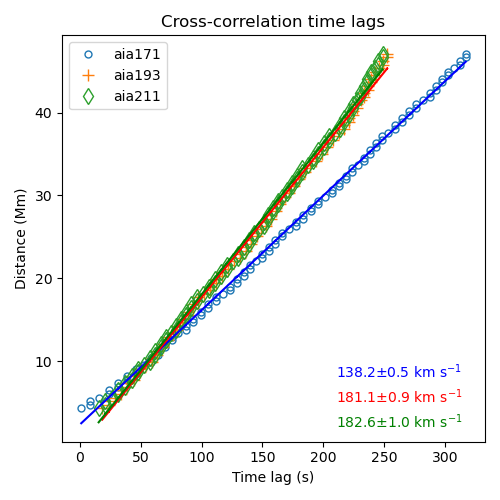} 
\caption{Time lags between the oscillations measured as a function of distance along the loop. Different symbols/colors denote data for different wavelength channels. The overplotted solid lines represent linear fits to the corresponding data. The propagation speeds estimated from the measured slopes are listed in the plot.}
\label{fig7}
\end{figure*}
The resulting time lags, as a function of distance, are shown for all three channels in Fig.{\,}\ref{fig7}. The inverse of the slope derived from a linear fit to the time lag values will give us the propagation speed. As listed in the figure, the speed values obtained are 138.2$\pm$0.5{\,}km{\,}s$^{-1}$, 181.1$\pm$0.9{\,}km{\,}s$^{-1}$, and 182.6$\pm$1.0{\,}km{\,}s$^{-1}$, respectively, for the 171{\,}{\AA}, 193{\,}{\AA}, and 211{\,}{\AA} channels. The expected propagation speeds according to their characteristic peak response temperatures are 135.1{\,}km{\,}s$^{-1}$, 180.5{\,}km{\,}s$^{-1}$, and 202.8{\,}km{\,}s$^{-1}$. Evidently, the obtained propagation speeds match pretty well with the expected speeds for the 171{\,}{\AA} and 193{\,}{\AA} channels but that in the 211{\,}{\AA} channel, surprisingly, does not match. Instead, the propagation speed of the 211{\,}{\AA} channel is nearly the same as that found in the 193{\,}{\AA}, possibly suggesting that the emission seen at this wavelength is close to the 193{\,}{\AA} peak temperature. Note that the line of sight angle considered here is perpendicular to the loop axis, so there is no projection effect unlike that found in observations. \\

\section{Summary and Conclusions}
\label{concl}
Employing a 3D MHD model, we study the propagation of slow magneto-acoustic waves in a multi-stranded, multi-thermal loop for the first time. The loop is modelled as a vertical cylinder composed of a bundle of 33 strands with randomly varying temperatures and densities. The gravitational stratification and other non-ideal effects are ignored. Slow waves are generated in the system by perturbing the vertical velocity at the bottom of the loop. The simulation results are then forward-modelled to produce synthetic images in all 6 coronal channels of SDO/AIA. Additionally, we estimate and add the corresponding data noise as would have been present during real observations. It is observed that the emission in the 193{\,}{\AA} channel is adjacent to that found in the 171{\,}{\AA} channel rather than being co-spatial. Although this behaviour is dependent on the distribution of plasma and the line of sight direction, we do note that the observed loops are known to exhibit similar traits. Time-distance maps are constructed from the synthetic data to study the propagation properties of slow waves. It is observed that the oscillations are only visible in certain channels and not visible in others and their appearance is related to the available plasma temperatures within the loop. The propagation speed of slow waves is also found to be sensitive to this information. Indeed, \citet{2024arXiv240109803V} have shown that the bandpass and the peak response temperature of an imaging filter can influence the wave propagation speeds that may be observed in a multi-thermal loop, in good agreement with the results found here.   \\

Observations in the past \citep[for e.g.,][]{2012ApJ...757..160J, 2012SoPh..279..427K, 2013ApJ...778...26U, 2017ApJ...834..103K, 2020A&A...638A...6S} have often shown that slow magneto-acoustic waves are visible only in certain wavelength channels. Additionally, their amplitudes and propagation speeds are found to vary from one channel to the other. While this differential propagation behaviour has been previously suggested to be due to a multi-thermal nature of the coronal loops \citep{2003A&A...404L...1K, 2017ApJ...834..103K}, our results demonstrate that we can take this to the next step and possibly infer the range of plasma temperatures within a coronal loop. This can be achieved, for example, by adjusting our loop model with an appropriate range of plasma temperatures that may reproduce the multi-wavelength behaviour of slow magneto-acoustic waves observed in a particular loop. The cross-field thermal information provides an important constraint for the coronal heating models so we believe this could be a significant step forward. The Differential Emission Measure (DEM) technique as well, in principle, provides this information but it suffers from line-of-sight integration effects. Importantly, it is highly non-trivial to discern whether the cross-field thermal characteristics obtained solely belong to the loop plasma or some background/foreground structures. On the other hand, because we use the properties of waves propagating within the loop here, one can stay assured that the thermal information obtained purely belongs to the loop.\\

In summary, via a multi-stranded loop model, we demonstrate that the seismological capabilities of slow magneto-acoustic waves can be further extended to obtain additional information on the cross-field thermal structure of coronal loops. We anticipate future studies to build on this and develop different plasma distribution models for comparison with the observation characteristics of slow waves. Moreover, because the slow waves are regularly observed in a variety of structures including active region fan loops, polar plumes, on-disk plumes, hot flare loops, etc., this technique is widely applicable in the solar corona and thus can be very useful once properly established.

\acknowledgements 
SKP is grateful to FWO Vlaanderen for a senior postdoctoral fellowship (No. 12ZF420N). The resources and services used in this work were provided by the VSC (Flemish Supercomputer Center), funded by the Research Foundation - Flanders (FWO) and the Flemish Government. SKP would like to thank Norbert Magyar for valuable discussions on AMRVAC simulations. TVD was supported by the European Research Council (ERC) under the European Union's Horizon 2020 research and innovation programme (grant agreement No 724326), the C1 grant TRACEspace of Internal Funds KU Leuven, and a Senior Research Project (G088021N) of the FWO Vlaanderen. TVD would like to thank Dipankar Banerjee, Vaibhav Pant and Krishna Prasad for their hospitality when visiting ARIES, Nainital in spring 2023, where this work was discussed. TVD received financial support from the Flemish Government under the long-term structural Methusalem funding program, project SOUL: Stellar evolution in full glory, grant METH/24/012 at KU Leuven, and also the DynaSun project (number 101131534 of HORIZON-MSCA-2022-SE-01).


\end{document}